\documentclass{article}
\usepackage{graphicx}

\evensidemargin=\oddsidemargin
\def\Title#1#2#3{%
    \baselineskip=18pt
    \begin{center}
          {\large\bf{#1} \\ }
          \bigskip\bigskip
          {#2} \\
          {#3} \\
    \end{center}}
\long\def\Abstract#1{%
         \bigskip
         \parbox{0.93\textwidth}{%
                 \begin{center}
                       {\bf Abstract} \\
                 \end{center}
                 \medskip{\baselineskip=14pt #1}
                 \vss}
         \bigskip}

\makeatletter
\renewcommand{\section}%
 {\@startsection{section}{1}{0pt}%
  {-3.25ex plus -1ex minus -.2ex}{1.5ex plus .2ex}%
  {\vspace*{5mm}\raggedright\large\bf }}
\renewcommand{\subsection}%
 {\@startsection{subsection}{2}{0pt}%
  {-2.25ex plus -.5ex minus -.2ex}{-1.5ex plus -.2ex}{\bf }}
\renewcommand{\subsubsection}%
 {\@startsection{subsubsection}{3}{0pt}%
  {-1.25ex plus -.2ex minus -.1ex}{-1.2ex plus -.2ex}{\bf }}

\makeatother

\begin{document}

\Title{On the appearance of time in the classical limit of quantum gravity}%
{R. I. Ayala O\~na\footnote{E-mail: {\tt ayyala@sfedu.ru}},
D. P. Kislyakova\footnote{E-mail: {\tt dkislyakova@sfedu.ru}} and
T. P. Shestakova\footnote{E-mail: {\tt shestakova@sfedu.ru}}}%
{Department of Theoretical and Computational Physics,
Southern Federal University,\\
Sorge St. 5, Rostov-on-Don 344090, Russia}

\Abstract{A possible solution of the problem of time in the Wheeler -- DeWitt quantum geometrodynamics is that time appears in semiclassical limit. Following this line of thinking, one can come to the Schrodinger equation for matter fields in curved spacetime with quantum-gravitational corrections. In the present paper, we study the semiclassical limit in the case of a closed isotropic model with a scalar field decomposed into modes. We analyse calculations made within frameworks of three approaches. The first approach was proposed by Kiefer and Singh. Since the Wheeler -- DeWitt equation does not contain a time derivative, it is constructed by means of a special mathematical procedure, time variable being a parameter along a classical trajectory of gravitational field. The second method was suggested in the paper of Maniccia and Montani who introduced the Kucha\v r -- Torre reference fluid as an origin of time. And the third is the extended phase space approach to quantization of gravity. In this approach, the temporal Schrodinger equation is argued to be more fundamental than the Wheeler -- DeWitt equation, and there is no problem of time. The origin of time is fixing of a reference frame of some observer, who can register macroscopic consequences of quantum gravitational phenomena in the Very Early Universe. To go to the semiclassical limit, the Born -- Oppenheimer approximation for gravity is used. In each of the approaches, in the order ${\cal O}(1/M)$, a temporal Schr\"odinger equation for matter fields in curved spacetime with quantum gravitational corrections is obtained. However, equations and corrections are different in various approaches, and the results depend on additional assumptions made within the scopes of these approaches.}

\section{Introduction}
Any viable physical theory must be verified by observations. Unfortunately, there are many approaches to quantum gravity, but not enough data to discriminate among them. The main source of our data about the Very Early Universe is cosmic microwave background radiation (CMB). In 2012, in the essay written for the Gravity Research Foundation essay competition \cite{KK1}, Kiefer and Kr\"amer noted that it would be of importance to make theoretical predictions of small quantum-gravitational effects that could be directly compared with observation data. They suggested to use the Born -- Oppenheimer approximation for a system of gravitational and matter fields, that implies division of the system into slow and fast changing parts. Gravity corresponds to the slow changing part, while matter fields are supposed to change fast.

One of the first authors who applied the Born -- Oppenheimer approximation to gravity was Padmanabhan \cite{Padma}. He used a toy quantum mechanical model describing two particles with masses $M$ and $m$, while $M\gg m$. The position of the heavy particle changes very slowly, and its motion can be described by a classical equation. One can neglect the effect of the light particle on the motion of the heavy one. The light particle follows the changes in the position of the heavy particle. Its behaviour is described from quantum mechanical point of view by a temporal Schr\"odinger equation taking into account the presence of the heavy particle in the background. This is the Born -- Oppenheimer approximation for molecules. The action is expanded in a series in the parameter $M^{-1}$, which plays the role of a coupling constant between the heavy and light particles.

If one makes analogy between the heavy particle and the gravitational field which is assumed to change slowly and between the light particle and matter fields which are assumed to change faster, one can believe that the toy model mimics gravity interacting with quantum matter fields. The gravitational field is a solution to the Einstein equations in empty space, while the back reaction of the matter fields being excluded. The wave function of the matter satisfies a Schr\"odinger equation with a Hamiltonian depending on the background gravitational field. Instead the parameter $M$, one can consider the constant multiplier before the gravitational action $\displaystyle\frac{c^3}{16\pi G}$ \cite{LL}, or any other parameter inversely proportional to $G$, for example, the Planckian mass squared, $m_{Pl}^2=\displaystyle\frac{\hbar c}G$. This is the Born -- Oppenheimer approximation for gravity. In the order ${\cal O}(M)$ one obtains an approximate Schr\"odinger equation, and in the next order, ${\cal O}(M^{-1})$, the Schr\"odinger equation with quantum gravitational corrections.

This ``semiclassical program'' was developed by many authors (see, for example, \cite{Singh,KS,Kiefer1,KK2,GMMN,MM,KP} and many others). Kiefer and his collaborators believe that this program also solves the well-known problem of time in the Wheeler -- DeWitt quantum geometrodynamics. They advocate the point of view that there had been no time in the Very Early Universe, in the realm of quantum gravity, and time had appeared only at the semiclassical stage of the Universe existence. In the approach of Kiefer and collaborators, time variable is introduced as a parameter along a classical trajectory of the system in its configurational space. Kiefer wrote in \cite{Kiefer2} that the emergence of the usual notion of spacetime within quantum cosmology needs an explanation. However, to explain something means to find out its cause. When we talk about some cause, we usually imply a sequence of events in time, and we get into a vicious circle. We have to give up our attempts to explain the appearance of time, or we should extend the notion of time to the realm of quantum gravity.

In other approaches, time appears as a result of fixing a reference frame. In \cite{MM}, the origin of time is introducing the Kucha\v r – Torre reference fluid. Introducing a reference frame is an essential part of the extended phase space approach \cite{SSV1,SSV2,SSV3,SSV4}, where the physical content of the Universe, that is, its geometry and matter fields spreading in space, is described from the viewpoint of an observer in a certain reference frame (see also \cite{Shest1}).

In fact, one can hardly name a physicist who would deal with quantum gravity and not write about the problem of time. In the framework of the Wheeler--DeWitt quantum geometrodynamics, let us mention the well-known and often-cited review papers by Kucha\v r \cite{Kuchar} and Isham \cite{Isham}. Among others are the papers \cite{SS1,SS2} and the recent work \cite{CK}. In the framework of supersymmetric quantum cosmology, we would like to refer to the paper \cite{GL,Moniz1} and especially to \cite{KLM,Moniz2}, where the semiclassical approximation is also discussed. Some aspects of the problem of time are touched in string cosmology \cite{Lidsey1,Lidsey2,GMV,GV,Gasperini}. The relation of the problem of time with statistical mechanics is analysed in \cite{Rovelli1,CJR}. The apologists of loop quantum gravity insist that the very notions of space and time must be modified \cite{Rovelli2}. This list can be continued.

In the present paper, we compare the results that are obtained in three different approaches in the semiclassical limit of a closed isotropic model of the Universe with a scalar field.  In Section 2, we discuss equations for a wave function of the Universe in the extended phase space approach and the Wheeler -- DeWitt geometrodynamics. In Section 3, we consider the semiclassical limit and the highest orders of expansion in a power series in the parameter $M$. In Section 4, we use various approaches to obtain a temporal Schr\"odinger equation for matter fields with gravitational field in background and analyse the differences. Quantum gravitational corrections are derived in Section 5, and conclusions are given in Section 6.

\section{The model, the Schr\"odinger equation and the Wheeler -- DeWitt equation}
We start from the effective action
\begin{equation}
\label{eff_act}
S_{eff}=S_{grav}+S_{mat}+S_{gf}+S_{ghost}.
\end{equation}
which is used in the path integral formulation of quantum field theory. Apart from gravitational and matter parts, $S_{grav}$ and $S_{mat}$, the action includes the gauge-fixing and ghost terms, $S_{gf}$ and $S_{ghost}$, which are gauge-noninvariant. As a rule, the path integral is considered under asymptotic boundary conditions that ensure its gauge invariance. Physically, the asymptotic boundary conditions correspond to initial and final (asymptotic) states with free particles. The situation in the theory of gravity is more complicated: Gravitating systems do not have asymptotic states, except asymptotically-flat spacetimes. Therefore, the asymptotic boundary conditions are not justified in the case of gravity.

One cannot derive a gauge-invariant Schr\"odinger equation from the path integral with the gauge-noninvariant effective action and without the asymptotic boundary conditions (see \cite{Shest2} for details). The Schr\"odinger equation turns out to be gauge-dependent.

We consider the system of gravitational field and a real scalar field with conformal coupling:
\begin{equation}
\label{gen_action}
S_{grav}+S_{mat}=\frac1{4\pi^2}\int\! d^4x\sqrt{-g}\left(\frac{c^3}{16\pi G}R
 +g_{\mu\nu}\partial^{\mu}\phi\partial^{\nu}\phi+\frac16 R\phi^2\right),
\end{equation}
where the coefficient $\displaystyle\frac1{4\pi^2}$ is introduced for convenience, and the coefficient
$\displaystyle\frac{c^3}{16\pi G}$ will be denoted below as $M$.

We shall use a closed isotropic model to illustrate our results. The spacetime interval looks like:
\begin{equation}
\label{interval}
ds^2=N^2(t)dt^2-a^2(t)\left[d\chi^2+\sin^2\chi\left(d\theta^2+\sin^2\theta d\varphi^2\right)\right].
\end{equation}
After redefinition $\phi=\sqrt{2}\pi\displaystyle\frac{\varphi}a$, the action for the scalar field reads:
\begin{equation}
\label{act_scal}
S_{mat}=\frac12\int d^4x\sqrt{\gamma}\left[\frac aN\dot{\varphi}^2
 -Na\varphi\varphi^{;i}_{;i}-\frac Na\varphi^2\right],
\end{equation}
where a dot denotes a time derivative, $\sqrt{\gamma}=\sin^2\chi\sin\theta$, $\varphi^{;i}_{;i}=-\Delta\varphi$ and
$\Delta$ is the Laplace -- Beltrami operator in a positive curvature space:
\begin{equation}
\label{LBO}
\Delta
 =\frac1{a^2\sin^2\chi}\left[\frac{\partial}{\partial\chi}\left(\sin^2\chi\frac{\partial}{\partial\chi}\right)
  +\frac1{\sin\theta}\frac{\partial}{\partial\theta}\left(\sin\theta\frac{\partial}{\partial\theta}\right) +\frac1{\sin^2\theta}\frac{\partial^2}{\partial\omega ^2}\right].
\end{equation}
Now, we present the scalar field as
\begin{equation}
\label{inhomo}
\varphi(t,\chi,\theta,\omega)=\sum_{n,l,m}f_{nlm}(t)\Phi_{nlm}(\chi,\theta,\omega),
\end{equation}
where $\Phi_{nlm}(\chi,\theta,\omega)$ are eigenfunctions of the Laplace -- Beltrami operator \cite{Grib},
\begin{equation}\label{Phi}
\Delta\Phi_{nlm}(\chi,\theta,\varphi)
 =\frac1{a^2}(1-\lambda_{nl}^2)\Phi_{nlm}(\chi,\theta,\varphi);\quad
\lambda_{nl}=1+n+l.
\end{equation}

After integrating over 3-space, the action (\ref{act_scal}) reads:
\begin{equation}
\label{act_scal_inhomo}
S_{mat}=\frac12\int dt\sum_{n,l,m}\left[\frac aN\dot f^2_{nlm}-\frac Na\lambda^2_{nl}f^2_{nlm}\right].
\end{equation}
Below we shall used indices ${\bf k}=(n,l,m)$. Therefore, effective action for our model is as following:
\begin{equation}
\label{model_eff_act}
S_{eff}
 =\int\! dt\left[\frac12\left(-M\frac{a\dot a^2}N+MNa\right)
  +\frac12\sum_{\bf k}\left(\frac aN\dot f^2_{\bf k}-\frac Na\lambda^2_{\bf k}f^2_{\bf k}\right)
  +\pi\left(\dot N-\frac{df}{da}\dot a\right)+\dot{\bar\theta}N\bar\theta\right].
\end{equation}
Here $\theta$, $\bar\theta$ are the Faddeev -- Popov ghosts; the gauge condition $N=f(a)+c$, $c=\rm{const}$ is used in a differential form, $\dot N=\displaystyle\frac{df}{da}\dot a$. It introduces the missing velocity $\dot N$ into the effective Lagrangian.

The method proposed by Feynman \cite{Feynman} and generalised by Cheng \cite{Cheng}, leads to the equation:
\begin{equation}
\label{Sch_eq_full}
i\hbar\frac{\partial\Psi}{\partial t}
 =-\frac{\hbar^2}{2M\mu}\frac{\partial}{\partial Q^{\alpha}}
   \left(\mu G^{\alpha\beta}\frac{\partial\Psi}{\partial Q^{\beta}}\right)
  -\frac M2 Na\Psi
  +\frac12\frac Na\sum_{\bf k}\left(-\hbar^2\frac{\partial^2\Psi}{\partial f^2_{\bf k}}
   +\lambda^2_{\bf k}f^2_{\bf k}\Psi\right)
  -\frac{\hbar^2}N\frac{\partial^2\Psi}{\partial\theta\partial\bar\theta}
\end{equation}
where $\mu$ is the measure in the path integral, for our model $\mu=\displaystyle\sqrt{\frac aN}$; $Q^{\alpha}=(N,a)$;
\begin{equation}
\label{Galbe_inv}
G^{\alpha\beta}=\left(
\begin{array}{cc}
-\displaystyle\frac Na\left(\frac{df}{da}\right)^2&-\displaystyle\frac Na\frac{df}{da}\\
-\displaystyle\frac Na\frac{df}{da}&-\displaystyle\frac Na
\end{array}
\right);
\end{equation}

The wave function $\Psi(N,a,f_{\bf k},\theta,\bar\theta;t)$ is defined on extended configurational space which includes physical as well as gauge and ghost variables. The general solution to the equation (\ref{Sch_eq_full}) is:
\begin{equation}
\label{GS}
\Psi(N,a,f_{\bf k},\theta,\bar\theta;t)
 =\int\Psi_c(a,f_{\bf k},t)\,\delta(N-f(a)-c)\,(\bar\theta+i\theta)\,dc.
\end{equation}

Of greatest interest is the Schr\"odinger equation for the physical part of the wave function $\Psi_c(a,f_{\bf k},t)$, whose explicit form for our model reads
\begin{equation}
\label{Sch_phys}
i\hbar\frac{\partial\Psi}{\partial t}
 =\frac{\hbar^2}{2M}\frac{f(a)}a\frac{\partial^2\Psi}{\partial a^2}
  -\frac{\hbar^2}{4M}\left(\frac{f(a)}{a^2}-\frac1a\frac{df}{da}\right)\frac{\partial\Psi}{\partial a}
  -\frac M2 f(a)a\Psi
  +\frac12\frac{f(a)}a\sum_{\bf k}\left(-\hbar^2\frac{\partial^2\Psi}{\partial f^2_{\bf k}}
   +\lambda^2_{\bf k}f^2_{\bf k}\Psi\right)
\end{equation}

The scalar field Hamiltonian operator in (\ref{Sch_phys}) resembles the one in \cite{KK2}. It is quadratic in momenta conjugate to the amplitudes $f_{\bf k}$. On the other hand, one can go from coordinate representation to the so-called holomorphic representation \cite{SF,Berez} by expressing the amplitudes $f_{\bf k}$ and their momenta $p_{\bf k}$ in terms of variables $b^*_{\bf k}$, $b_{\bf k}$:
\begin{equation}
\label{holomorph}
f_{\bf k}=\sqrt{\frac{\hbar}{2\lambda_{\bf k}}}\left(b^*_{\bf k}+b_{\bf k}\right)\quad
p_{\bf k}=i\sqrt{\frac{\hbar\lambda_{\bf k}}2}\left(b^*_{\bf k}-b_{\bf k}\right).
\end{equation}

The action (\ref{model_eff_act}) written in terms of $b^*_{\bf k}$, $b_{\bf k}$ reads:
\begin{equation}
\label{eff_act_holom}
S_{eff}
 =\int\! dt\left[\frac12\left(-M\frac{a\dot a^2}N+MNa\right)
  -\frac12\sum_{\bf k}\left(ib^*_{\bf k}\dot b_{\bf k}
   +\frac Na\hbar\lambda_{\bf k}b^*_{\bf k}b_{\bf k}\right)
  +\pi\left(\dot N-\frac{df}{da}\dot a\right)+\dot{\bar\theta}N\bar\theta\right].
\end{equation}
And the Schr\"odinger equation for the physical part of the wave function $\Psi(a,b^*_{\bf k},t)$ that follows from this action looks like
\begin{equation}
\label{Sch_phys_holom}
i\hbar\frac{\partial\Psi}{\partial t}
 =\frac{\hbar^2}{2M}\frac{f(a)}a\frac{\partial^2\Psi}{\partial a^2}
  -\frac{\hbar^2}{4M}\left(\frac{f(a)}{a^2}-\frac1a\frac{df}{da}\right)\frac{\partial\Psi}{\partial a}
  -\frac M2 f(a)a\Psi
  +\frac{\hbar}2\frac{f(a)}a\sum_{\bf k}\lambda_{\bf k}b^*_{\bf k}\frac{\partial\Psi}{\partial b^*_{\bf k}}.
\end{equation}

One can check that the Wheeler -- DeWitt equation for the model could be obtained from the Schr\"odinger equation for the physical part of the wave function under the conditions
$\displaystyle\frac{\partial\Psi}{\partial t}=0$ and $N=1$ (or $f(a)=1$). Therefore, instead of (\ref{Sch_phys}) and (\ref{Sch_phys_holom}) one gets the following equations:
\begin{equation}
\label{WDW}
0=\frac{\hbar^2}{2Ma}\frac{\partial^2\Psi}{\partial a^2}
  -\frac{\hbar^2}{4Ma^2}\frac{\partial\Psi}{\partial a}
  -\frac M2 a\Psi
  +\frac1{2a}\sum_{\bf k}\left(-\hbar^2\frac{\partial^2\Psi}{\partial f^2_{\bf k}}
   +\lambda^2_{\bf k}f^2_{\bf k}\Psi\right);
\end{equation}
\begin{equation}
\label{WDW_holom}
0=\frac{\hbar^2}{2Ma}\frac{\partial^2\Psi}{\partial a^2}
  -\frac{\hbar^2}{4Ma^2}\frac{\partial\Psi}{\partial a}
  -\frac M2 a\Psi
  +\frac{\hbar}{2a}\sum_{\bf k}\lambda_{\bf k}b^*_{\bf k}\frac{\partial\Psi}{\partial b^*_{\bf k}}.
\end{equation}

\section{The semiclassical limit and the highest orders of expansion}
Going to the semiclassical limit, we present the wave function in the form
\begin{equation}
\label{WKB_WF}
\Psi=\exp\left(\frac{iS}{\hbar}\right),
\end{equation}
and then we expand $S$ as
\begin{equation}
\label{BO_appr}
S=MS_0+S_1+\frac1M S_2+{\cal O}\left(\frac1{M^2}\right).
\end{equation}
Then, we substitute (\ref{WKB_WF}) and (\ref{BO_appr}) into the equations (\ref{Sch_phys}) and (\ref{WDW}). The highest order, ${\cal O}(M^2)$, yields:
\begin{equation}
\label{M2}
\sum_{\bf k}\left(\frac{\partial S_0}{\partial f_{\bf k}}\right)^2=0.
\end{equation}
It means that $S_0$ does not depend on the scalar field modes, but only on the scale factor $a$. Therefore, the gravitational field is separated from matter. In the next order, ${\cal O}(M)$, from (\ref{WDW}), one gets the Hamilton -- Jacobi equation for pure gravity:
\begin{equation}
\label{M1}
\frac1a\left(\frac{\partial S_0}{\partial a}\right)^2+a=0.
\end{equation}
In the extended phase space approach, one obtains from (\ref{Sch_phys}):
\begin{equation}
\label{M1_EPS}
-\frac{\partial S_0}{\partial t}
 =-\frac12\frac{f(a)}a\left(\frac{\partial S_0}{\partial a}\right)^2-\frac12 f(a)a=0.
\end{equation}
To be in agreement with the classical Einstein theory, one should put
$\displaystyle\frac{\partial S_0}{\partial t}=0$. Then, the gauge-fixing function $f(a)$ can be cancelled out and one again comes to the Hamilton -- Jacobi equation (\ref{M1}). Its solution is
\begin{equation}
\label{S0}
S_0=i\frac{a^2}2.
\end{equation}
Let us note that in \cite{KS} $S_{0}$ is assumed to be real. But, as one can see, in the case of a closed Universe, $S_{0}$ is complex.

In the case of holomorphic representation of scalar field, after substitution of (\ref{WKB_WF}) and (\ref{BO_appr}) into (\ref{Sch_phys_holom}) and (\ref{WDW_holom}), one would not obtain any terms of the order ${\cal O}(M^2)$, and gravity cannot be separated from the scalar field. One needs some additional assumptions, for example, that $S_0=S_0(a)$. Obviously, this requirement is equivalent to (\ref{M2}), but now it does not immediately result from the expansion (\ref{BO_appr}). From now on, we shall consider only the equations (\ref{Sch_phys_holom}) and (\ref{WDW_holom}) with the scalar field in holomorphic representation.

In \cite{GMMN,MM} the assumption is made that, for our model, reads as:
\begin{equation}
\label{MM_ass}
S_0=\sigma_0(a),\quad
S_1=\sigma_1(a)+\eta_1(a,b^*_{\bf k}),\quad
S_2=\sigma_2(a)+\eta_2(a,b^*_{\bf k}),
\end{equation}
i.e. the wave function (\ref{WKB_WF}) can be written as:
\begin{equation}
\label{MM_WF}
\Psi=\exp\left[\frac i{\hbar}\left(M\sigma_0+\sigma_1+\frac1M\sigma_2\right)\right]
     \exp\left[\frac i{\hbar}\left(\eta_1+\frac1M\eta_2\right)\right].
\end{equation}
Thus, the wave function is presented as a product of gravitational and matter wave functions. According to another assumption accepted in \cite{MM}, the energy of matter fields is much smaller with respect to the energy of gravity, and the gravitational wave function must satisfy constraint equations for pure gravity. In this approach, one obtains equations for pure gravity in each order of expansion, but not only in the order ${\cal O}(M)$, where we have got the Hamilton -- Jacobi equation (\ref{M1}). This simplifies the equations for the system ``gravity + matter'' and leads to different results than in the approach by Kiefer and his collaborators \cite{KS,KK1,KK2,KP}.

\section{Derivation of the temporal Schr\"odinger equation}
In the next order, ${\cal O}(M^0)$, one would like to get a temporal Schr\"odinger equation for the scalar field with the gravitational field in the background. We shall start from the Wheeler -- DeWitt geometrodynamics, since, as it is well-known, time is absent in this approach. Below we follow the method described in \cite{KS}. So, in this order, one obtains from (\ref{WDW_holom}) the equation:
\begin{equation}
\label{M0_WDW}
0=\frac{i\hbar}{2a}\frac{\partial^2 S_0}{\partial a^2}
 -\frac1a\frac{\partial S_0}{\partial a}\frac{\partial S_1}{\partial a}
 -\frac{i\hbar}{4a^2}\frac{\partial S_0}{\partial a}
 +\frac i{2a}\sum_{\bf k}\lambda_{\bf k}b^*_{\bf k}\frac{\partial S_1}{\partial b^*_{\bf k}}.
\end{equation}

There is no time derivative in this equation, but it can be constructed, if, firstly, we define the wave function according to
\begin{equation}
\label{chi_fun}
\chi(a,\phi)=D(a)\exp\left(\frac i{\hbar}S_1\right),
\end{equation}
where $D(a)$ is an unknown function. Secondly, we note that operating on (\ref{chi_fun}) with the Hamiltonian of the scalar field,
\begin{equation}
\label{Ham_mat}
H_m=\frac{\hbar}{2a}\sum_{\bf k}\lambda_{\bf k}b^*_{\bf k}\frac{\partial}{\partial b^*_{\bf k}},
\end{equation}
we would obtain the last term in (\ref{M0_WDW}) multiplied by $\chi$. After multiplication (\ref{M0_WDW}) by $\chi$ it can be rewritten as
\begin{equation}
\label{M0_WDW1}
-\frac{i\hbar}{2a}\frac{\partial^2 S_0}{\partial a^2}\chi
 +\frac1a\frac{\partial S_0}{\partial a}\frac{\partial S_1}{\partial a}\chi
 +\frac{i\hbar}{4a^2}\frac{\partial S_0}{\partial a}\chi
 =H_m\chi.
\end{equation}
Therefore, to gain a required Schr\"odinger equation, one should somehow replace the three terms in the left-hand side of (\ref{M0_WDW1}) by a term with a time derivative of the wave function. Since $S_0$ is the classical action for gravity, its derivative $\displaystyle\frac{\partial S_0}{\partial a}$ is a tangent vector to a classical trajectory of the gravitational field in its configurational space. A projection of gradient of the function $\chi$ on the direction of the tangent vector gives a derivative of $\chi$ with respect to a parameter $\tau$ along a classical trajectory that plays a part of a time variable in this approach. Indeed, in a case of configurational space with coordinates $q^a$, denoting as $p_a$ their conjugate momenta, we have:
\begin{equation}
\label{time_oper_gen1}
\frac{\partial\chi}{\partial\tau}
 =\frac{\partial\chi}{\partial q^a}\frac{\partial q^a}{\partial\tau}
 =G^{ab}p_b\frac{\partial\chi}{\partial q^a}
 =G^{ab}\frac{\partial S}{\partial q^b}\frac{\partial\chi}{\partial q^a}
 =({\bf p},\nabla)\chi,
\end{equation}
where $G^{ab}$ is the inverse configurational space metric. In our model, there is only one physical degree of freedom of the gravitational field, the scale factor $a$, and the role of $G^{ab}$ is played by the inverse multiplier at the generalised velocity $\dot a^2$ in the action (\ref{eff_act_holom}),
$-\displaystyle\frac Na$, the minus sign being a feature of gravity. With $N=1$ it results in
\begin{equation}
\label{time_oper}
\frac{\partial\chi}{\partial\tau}
 =-\frac1a\frac{\partial S}{\partial a}\frac{\partial\chi}{\partial a}.
\end{equation}

Operating on (\ref{chi_fun}) with the operator in the right-hand side of (\ref{time_oper}), we get:
\begin{equation}
\label{time_oper_chi}
-\frac{i\hbar}a\frac{\partial S_0}{\partial a}\frac{\partial\chi}{\partial a}
 =-\frac{i\hbar}a\frac{\partial S_0}{\partial a}\frac1D\frac{dD}{da}\chi
  +\frac1a\frac{\partial S_0}{\partial a}\frac{\partial S_1}{\partial a}\chi.
\end{equation}
Equating the left-hand side of (\ref{M0_WDW1}) to the right-hand side of (\ref{time_oper_chi}), one obtains an equation for $D(a)$:
\begin{equation}
\label{D_eq}
\frac1{2a}\frac{\partial^2 S_0}{\partial a^2}
 -\frac1{4a^2}\frac{\partial S_0}{\partial a}
 =\frac1a\frac{\partial S_0}{\partial a}\frac1D\frac{dD}{da}.
\end{equation}

It can be solved taking into account the solution (\ref{S0}) to the Hamilton -- Jacobi equation and yields:
\begin{equation}
\label{D_a}
D(a)=a^{\frac14}.
\end{equation}

Finally, we have come to the temporal Schrodinger equation:
\begin{equation}
\label{temp_SE}
i\hbar\frac{\partial\chi}{\partial\tau}=H_m\chi.
\end{equation}

Obviously, these calculations cannot give us an answer to the question, how time had emerged from the timeless Very Early Universe. Moreover, here the introduction of time is based on the supposition that a classical spacetime had already existed. The appearance of time remains to be a mystery without any key.

Introducing the function $D(a)$ is not presumed in the WKB or Born -- Oppenheimer approximation. However, it would be impossible to obtain a time derivative without this little mathematical artifice.

In other approaches, the time derivative appears in equations of quantum geometrodynamics as a consequence of introducing a reference frame. In \cite{MM}, the origin of time is introducing the Kucha\v r -- Torre reference fluid \cite{KT}. It results in additional terms that can be combined to define a time derivative, which appears in the left-hand side of (\ref{WDW_holom}). Since the authors accept the ideology of the Wheeler -- DeWitt theory, gravitational part of the action, i.e. $\sigma_0$, $\sigma_1$, $\sigma_2$ does not depend on a time variable. In the order ${\cal O}(M^0)$, this approach yields the following equations for pure gravity and for the whole system, respectively:
\begin{eqnarray}
\label{M0_MM_grav}
0&=&\frac{i\hbar}{2a}\frac{d^2\sigma_0}{da^2}
 -\frac1a\frac{d\sigma_0}{da}\frac{d\sigma_1}{da}
 -\frac{i\hbar}{4a^2}\frac{d\sigma_0}{da};\\
\label{M0_MM}
-\frac{\partial\eta_1}{\partial t}
&=&\frac{i\hbar}{2a}\frac{d^2\sigma_0}{da^2}
 -\frac1a\frac{d\sigma_0}{da}\frac{d\sigma_1}{da}
 -\frac1a\frac{d\sigma_0}{da}\frac{\partial\eta_1}{\partial a}
 -\frac{i\hbar}{4a^2}\frac{d\sigma_0}{d a}
 +\frac i{2a}\sum_{\bf k}\lambda_{\bf k}b^*_{\bf k}\frac{\partial\eta_1}{\partial b^*_{\bf k}}.
\end{eqnarray}
Due to (\ref{M0_MM_grav}), three terms in (\ref{M0_MM}) disappear, and one comes to the equation:
\begin{equation}
\label{M0_MM_red}
-\frac{\partial\eta_1}{\partial t}
=-\frac1a\frac{d\sigma_0}{da}\frac{\partial\eta_1}{\partial a}
 +\frac i{2a}\sum_{\bf k}\lambda_{\bf k}b^*_{\bf k}\frac{\partial\eta_1}{\partial b^*_{\bf k}}.
\end{equation}
Defining the function $\chi$ as
\begin{equation}
\label{chi_MM}
\chi(a,\phi)=\exp\left(\frac i{\hbar}\eta_1\right)
\end{equation}
and making use of the operator (\ref{Ham_mat}), we rewrite (\ref{M0_MM_red}) as
\begin{equation}
\label{Sch_MM}
i\hbar\frac{\partial\chi}{\partial\tau}
 =H_m\chi+\frac{i\hbar}a\frac{d\sigma_0}{da}\frac{\partial\chi}{\partial a}.
\end{equation}
The last term in the right-hand side of (\ref{Sch_MM}) looks redundant. The authors of \cite{MM} noted that an operator ordering can be chosen so that the term with the first derivative of the wave function with respect to $a$ vanishes. However, discussing an open isotropic model, they have not defined the required ordering.

In the extended phase space approach, the operator ordering is defined by the procedure of derivation of the Schr\"odinger equation from the path integral. It is a temporal Schr\"odinger equation, and we do not need to construct the time derivative, we already have it in the equation. In the order ${\cal O}(M^0)$, from Eq. (\ref{Sch_phys_holom}) we get:
\begin{equation}
\label{M0_EPS}
-\frac{\partial S_1}{\partial t}
 =\frac{i\hbar}2\frac{f(a)}a\frac{\partial^2 S_0}{\partial a^2}
  -\frac{f(a)}a\frac{\partial S_0}{\partial a}\frac{\partial S_1}{\partial a}
  -\frac{i\hbar}4\left(\frac{f(a)}{a^2}-\frac1a\frac{df}{da}\right)\frac{\partial S_0}{\partial a}
  +\frac i2\frac{f(a)}a\sum_{\bf k}\lambda_{\bf k}b^*_{\bf k}\frac{\partial S_1}{\partial b^*_{\bf k}}.
\end{equation}

Now we define the Hamiltonian of the scalar field as
\begin{equation}
\label{Ham_mat_EPS}
H_{mat}=\frac{\hbar}2\frac{f(a)}a\sum_{\bf k}\lambda_{\bf k}b^*_{\bf k}\frac{\partial}{\partial b^*_{\bf k}},
\end{equation}
it depends on the gauge fixing function $f(a)$ and coincides with (\ref{Ham_mat}) when $f(a)=1$. To obtain the Schr\"odinger equation in the form
\begin{equation}
\label{temp_SE_EPS}
i\hbar\frac{\partial\chi}{\partial t}=H_{mat}\chi
\end{equation}
one should require that the first three terms in the right-hand side of (\ref{M0_EPS}) would be equal to zero:
\begin{equation}
\label{S1}
\frac{i\hbar}2\frac{f(a)}a\frac{\partial^2 S_0}{\partial a^2}
 -\frac{f(a)}a\frac{\partial S_0}{\partial a}\frac{\partial S_1}{\partial a}
 -\frac{i\hbar}4\left(\frac{f(a)}{a^2}-\frac1a\frac{df}{da}\right)\frac{\partial S_0}{\partial a}=0.
\end{equation}

One can consider this as an equation for $S_1$. For example, if $f(a)=1$, one immediately gets
$S_1=\displaystyle\frac{i\hbar}4\ln a$. However, we now do not have an equation for $D(a)$.

\section{Quantum gravitational corrections to the Schr\"odinger equation}
We now turn to the order ${\cal O}(M^{-1})$. From the Wheeler -- DeWitt equation (\ref{WDW_holom}), one obtains:
\begin{equation}
\label{M-1}
0=\frac{i\hbar}{2a}\frac{\partial^2S_1}{\partial a^2}
 -\frac1{2a}\left(\frac{\partial S_1}{\partial a}\right)^2
 -\frac1a\frac{\partial S_0}{\partial a}\frac{\partial S_2}{\partial a}
 -\frac{i\hbar}{4a^2}\frac{\partial S_1}{\partial a}
 +\frac i{2a}\sum_{\bf k}\lambda_{\bf k}b^*_{\bf k}\frac{\partial S_2}{\partial b^*_{\bf k}}.
\end{equation}
Returning to the function (\ref{chi_fun}), we can express the derivatives of $S_1$ with respect to $a$ and substitute then into (\ref{M-1}):
\begin{eqnarray}
\label{M-1-D}
0&=&\frac{\hbar^2}{2a}\left[\frac1{\chi}\frac{\partial^2\chi}{\partial a^2}
   -\frac2{D\chi}\frac{dD}{da}\frac{\partial\chi}{\partial a}
   -\frac1D\frac{d^2D}{da^2}
   +\frac2{D^2}\left(\frac{dD}{da}\right)^2\right]
  -\frac1a\frac{\partial S_0}{\partial a}\frac{\partial S_2}{\partial a}\nonumber\\
&-&\frac{\hbar^2}{4a^2}\left(\frac1{\chi}\frac{\partial\chi}{\partial a}-\frac1D\frac{dD}{da}\right)
  +\frac i{2a}\sum_{\bf k}\lambda_{\bf k}b^*_{\bf k}\frac{\partial S_2}{\partial b^*_{\bf k}}.
\end{eqnarray}

The solution for $S_2$ is sought in the form $S_2=\sigma_2(a)+\eta_2(a,b^*_{\bf k})$ (compare with (\ref{MM_ass})), where $\sigma_2(a)$ is a solution to the equation
\begin{equation}
\label{sigma2}
\frac1a\frac{\partial S_0}{\partial a}\frac{d\sigma_2}{da}
 -\frac{\hbar^2}{4Da^2}\frac{dD}{da}
 +\frac{\hbar^2}{2a}\left[\frac1D\frac{d^2D}{da^2}-\frac2{D^2}\left(\frac{dD}{da}\right)^2\right]=0.
\end{equation}
Thus, we separate our all terms in the equation (\ref{sigma2}) depending only on $a$. When $D(a)=a^{\frac14}$ (see (\ref{D_a})), this equation is reduced to the form:
\begin{equation}
\label{sigma2a}
i\frac{d\sigma_2}{da}-\frac{7\hbar^2}{32a^3}=0.
\end{equation}
One can check that exactly the same equation for $\sigma_2$ is obtained in the second order,
${\cal O}(\hbar^2)$, of the WKB expansion for the gravitational part of the wave function.

The rest terms in (\ref{M-1-D}) are included into the equation for the function $\eta_2$:
\begin{equation}
\label{eta2}
\frac1a\frac{\partial S_0}{\partial a}\frac{\partial\eta_2}{\partial a}
 =\frac{\hbar^2}{2a}\left(\frac1{\chi}\frac{\partial^2\chi}{\partial a^2}
 -\frac2{D\chi}\frac{dD}{da}\frac{\partial\chi}{\partial a}\right)
 -\frac{\hbar^2}{4a^2\chi}\frac{\partial\chi}{\partial a}
 +\frac i{2a}\sum_{\bf k}\lambda_{\bf k}b^*_{\bf k}\frac{\partial\eta_2}{\partial b^*_{\bf k}}.
\end{equation}

We introduce a new function
\begin{equation}
\label{xi_fun}
\xi=\chi\exp\left(\frac{i\eta_2}{\hbar M}\right).
\end{equation}

Let us note that
\begin{equation}
\label{H_M_xi}
H_m\xi=\frac{\xi}{\chi}H_m\chi
 +\frac i{2Ma}\xi\sum_{\bf k}\lambda_{\bf k}b^*_{\bf k}\frac{\partial\eta_2}{\partial b^*_{\bf k}};
\end{equation}
\begin{equation}
\label{t-d-tr}
\frac1a\frac{\partial S_0}{\partial a}\frac{\partial\eta_2}{\partial a}
 =-i\hbar M\frac1a\frac{\partial S_0}{\partial a}\left(\frac1{\xi}\frac{\partial\xi}{\partial a}
  -\frac1{\chi}\frac{\partial\chi}{\partial a}\right)
 =i\hbar M\left(\frac1{\xi}\frac{\partial\xi}{\partial\tau}
  -\frac1{\chi}\frac{\partial\chi}{\partial\tau}\right).
\end{equation}
Dividing (\ref{eta2}) by $M$, multiplying by $\xi$ and making use of (\ref{H_M_xi}) and (\ref{t-d-tr}), one comes to the equation
\begin{equation}
\label{xi_corr_D}
i\hbar\frac{\partial\xi}{\partial\tau}
 =H_m\xi
  +\frac{\hbar^2}{2Ma}\xi\left(\frac1{\chi}\frac{\partial^2\chi}{\partial a^2}
   -\frac2{D\chi}\frac{dD}{da}\frac{\partial\chi}{\partial a}\right)
  -\frac{\hbar^2}{4Ma^2}\frac{\xi}{\chi}\frac{\partial\chi}{\partial a},
\end{equation}
or, with $D(a)=a^{\frac14}$, this equation would be
\begin{equation}
\label{xi_corr}
i\hbar\frac{\partial\xi}{\partial\tau}
 =H_m\xi
  +\frac{\hbar^2}{2Ma\chi}\xi\left(\frac{\partial^2\chi}{\partial a^2}
   -\frac1a\frac{\partial\chi}{\partial a}\right).
\end{equation}

Taking into account that
\begin{equation}
\label{der-chi-xi}
\frac{\partial\chi}{\partial a}
 =\frac{\partial\xi}{\partial a}\frac{\chi}{\xi}+{\cal O}\left(\frac1M\right);
\quad
\frac{\partial^2\chi}{\partial a^2}
 =\frac{\partial^2\xi}{\partial a^2}\frac{\chi}{\xi}+{\cal O}\left(\frac1M\right),
\end{equation}
we finally obtain:
\begin{equation}
\label{xi-fin}
i\hbar\frac{\partial\xi}{\partial\tau}
 =H_m\xi
 +\frac{\hbar^2}{2Ma}\left(\frac{\partial^2\xi}{\partial a^2}
  -\frac1a\frac{\partial\xi}{\partial a}\right).
\end{equation}

This is the required Schr\"odinger equation with quantum gravitational corrections obtained in the framework of the approach used by Kiefer and his collaborations \cite{KS}.

In the approach described in \cite{MM}, in the order ${\cal O}(M^{-1})$, one has the equations for pure gravity and for the whole system:
\begin{eqnarray}
\label{M-1_MM_grav}
0&=&\frac{i\hbar}{2a}\frac{d^2\sigma_1}{da^2}
 -\frac1{2a}\left(\frac{d\sigma_1}{da}\right)^2
 -\frac1a\frac{d\sigma_0}{da}\frac{d\sigma_2}{da}
 -\frac{i\hbar}{4a^2}\frac{d\sigma_1}{da};\\
\label{M-1_MM}
-\frac{\partial\eta_2}{\partial t}
&=&\frac{i\hbar}{2a}\left(\frac{d^2\sigma_1}{da^2}
   +\frac{\partial^2\eta_1}{\partial a^2}\right)
 -\frac1{2a}\left[\left(\frac{d\sigma_1}{da}\right)^2
   +2\frac{d\sigma_1}{da}\frac{\partial\eta_1}{\partial a}
   +\left(\frac{\partial\eta_1}{\partial a}\right)^2\right]\nonumber\\
&-&\frac1a\frac{d\sigma_0}{da}\left(\frac{d\sigma_2}{da}
   +\frac{\partial\eta_2}{\partial a}\right)
 -\frac{i\hbar}{4a^2}\left(\frac{d\sigma_1}{da}
   +\frac{\partial\eta_1}{\partial a}\right)
 +\frac i{2a}\sum_{\bf k}\lambda_{\bf k}b^*_{\bf k}\frac{\partial\eta_2}{\partial b^*_{\bf k}}.
\end{eqnarray}
Omitting pure gravitational terms due to (\ref{M-1_MM_grav}), the following equation can be written:
\begin{equation}
\label{M-1_MM_red}
-\frac{\partial\eta_2}{\partial t}
=\frac{i\hbar}{2a}\frac{\partial^2\eta_1}{\partial a^2}
 -\frac1a\frac{d\sigma_1}{da}\frac{\partial\eta_1}{\partial a}
 -\frac1{2a}\left(\frac{\partial\eta_1}{\partial a}\right)^2
 -\frac1a\frac{d\sigma_0}{da}\frac{\partial\eta_2}{\partial a}
 -\frac{i\hbar}{4a^2}\frac{\partial\eta_1}{\partial a}
 +\frac i{2a}\sum_{\bf k}\lambda_{\bf k}b^*_{\bf k}\frac{\partial\eta_2}{\partial b^*_{\bf k}}.
\end{equation}

Further, one defines the function
\begin{equation}
\label{xi_fun_MM}
\xi=\exp{\left[\frac i{\hbar}\left(\eta_1+\frac1M\eta_2\right)\right]},
\end{equation}
which coincides with (\ref{xi_fun}) if $\chi$ is given by (\ref{chi_MM}). On the next step, dividing Eq. (\ref{M-1_MM_red}) by $M$ and summing it with Eq. (\ref{M0_MM_red}), we get the equation:
\begin{eqnarray}
\label{xi_eq1}
-\frac{\partial}{\partial t}\left(\eta_1+\frac1M\eta_2\right)
&=&-\frac1a\frac{d\sigma_0}{da}\frac{\partial}{\partial a}
   \left(\eta_1+\frac1M\eta_2\right)
  +\frac{i\hbar}{2Ma}\left[\frac{\partial^2\eta_1}{\partial a^2}
   +\frac i{\hbar}\left(\frac{\partial\eta_1}{\partial a}\right)^2\right]\nonumber\\
&-&\frac1{Ma}\frac{d\sigma_1}{da}\frac{\partial\eta _1}{\partial a}
  -\frac{i\hbar}{4Ma^2}\frac{\partial\eta_1}{\partial a}
  +\frac i{2a}\sum_{\bf k}\lambda_{\bf k}b^*_{\bf k}\frac{\partial}{\partial b^*_{\bf k}}
   \left(\eta_1+\frac1M\eta_2\right).
\end{eqnarray}
Expressing the derivatives of $\eta_1$ through derivatives of $\chi$ according to (\ref{chi_MM}), and using again (\ref{der-chi-xi}), we finally come to the equation for the function $\xi$:
\begin{equation}
\label{xi_eq}
i\hbar\frac{\partial\xi}{\partial t}
 =H_m\xi+\frac{i\hbar}a\frac{d\sigma_0}{da}\frac{\partial\xi}{\partial a}
  +\frac{\hbar^2}{2Ma}\frac{\partial^2\xi}{\partial a^2}
  +\frac{i\hbar}{Ma}\frac{d\sigma_1}{da}\frac{\partial\xi}{\partial a}
  -\frac{\hbar^2}{4Ma^2}\frac{\partial\xi}{\partial a}.
\end{equation}

In \cite{MM}, the authors have made another assumption that the derivatives of the matter part of the wave function with respect to slowly changing gravitational variables are small. It means that
\begin{equation}
\label{deriv_MM}
\frac{\partial\chi}{\partial a}={\cal O}\left(\frac1M\right),
\end{equation}
and the last three terms in the right-hand side of (\ref{xi_eq}) turn out to be of the order
${\cal O}\left(\displaystyle\frac1{M^2}\right)$ and should be omitted.

On the other hand, one can find $\sigma_1$ from (\ref{M0_MM_grav}), (\ref{S0}) and (\ref{MM_ass}):
\begin{equation}
\label{sigma1}
\sigma_1=\frac{i\hbar}4\ln a.
\end{equation}
Let us note that the equations (\ref{M0_MM_grav}) and (\ref{M-1_MM_grav}) are obtained in the first and second orders, respectively, of the WKB expansion for the gravitational part of the wave function. Substitution of (\ref{S0}) and (\ref{sigma1}) into (\ref{M-1_MM_grav}) yields Eq. (\ref{sigma2a}). If one substitutes (\ref{sigma1}) into (\ref{xi_eq}), the equation for $\xi$ would look as
\begin{equation}
\label{xi_eq2}
i\hbar\frac{\partial\xi}{\partial t}
 =H_m\xi+\frac{i\hbar}a\frac{d\sigma_0}{da}\frac{\partial\xi}{\partial a}
  +\frac{\hbar^2}{2Ma}\left(\frac{\partial^2\xi}{\partial a^2}
  -\frac1a\frac{\partial\xi}{\partial a}\right).
\end{equation}
Obviously, the last terms are the same corrections that were obtained above in the framework of the approach of Kiefer and his collaborations (see (\ref{xi-fin})), but because the assumption (\ref{deriv_MM}) they should be expunged. This is a real contradiction between the two approaches.

Then, the only correction is the second term in the right-hand side of (\ref{xi_eq2}). In the approach of Kiefer and collaborators this correction cannot appear, since the term gives rise to the time derivative in (\ref{xi-fin}). In the opinion of Maniccia and Montani \cite{MM}, the advantage of their approach is that this quantum gravitational correction is Hermitian. They argue (see also \cite{GMMN}) that $\sigma_0$ is a classical action, i.e. a solution to the Hamilton -- Jacobi equation and must be real. However, it is not true for our simple model. Should we believe that it is a defect of the model? Which is more important, in their notation, the operator acting on the function $\xi$ looks like $(-i\hbar\nabla_g)$, i.e. the operators of conjugate momenta of gravitational variables. It implicitly assumes that the measure in the inner product is trivial; otherwise the operator would not be Hermitian. But in our model the operator is $\displaystyle\frac{i\hbar}a\frac{\partial}{\partial a}$, and the measure
$\mu=\sqrt{\displaystyle\frac aN}$, it is determined by the procedure of derivation of the Schr\"odinger equation from the path integral. The general hermiticity condition for an operator $\hat A$ reads:
\begin{equation}
\label{herm}
\int\Psi^*(q)\left(\hat A\Phi(q)\right)\mu(q)dq
=\int\left(\hat A^*\Psi^*(q)\right)\Phi(q)\mu(q)dq,
\end{equation}
where $\Psi(q)$, $\Phi(q)$ are wave functions in the coordinate representation. One can check that the operator in the second term in the right-hand side of (\ref{xi_eq2}) does not satisfy the hermiticity condition.

At last, we shall discuss the extended phase space approach. We start from the equation
\begin{equation}
\label{M-1_EPS}
-\frac{\partial S_2}{\partial t}
 =i\hbar\frac{f(a)}{2a}\frac{\partial^2 S_1}{\partial a^2}
  -\frac{f(a)}{2a}\left(\frac{\partial S_1}{\partial a}\right)^2
  -\frac{f(a)}a\frac{\partial S_0}{\partial a}\frac{\partial S_2}{\partial a}
  -\frac{i\hbar}4\left(\frac{f(a)}{a^2}-\frac1a\frac{df}{da}\right)\frac{\partial S_1}{\partial a}
  +\frac i2\frac{f(a)}a\sum_{\bf k}\lambda_{\bf k}b^*_{\bf k}\frac{\partial S_2}{\partial b^*_{\bf k}},
\end{equation}
that follows from (\ref{Sch_phys_holom}) in the order ${\cal O}(M^{-1})$. We write down the analogue of (\ref{M-1-D}):
\begin{eqnarray}
\label{M-1-D-EPS}
-\frac{\partial S_2}{\partial t}
&=&\hbar^2\frac{f(a)}{2a}\left[\frac1{\chi}\frac{\partial^2\chi}{\partial a^2}
   -\frac2{D\chi}\frac{dD}{da}\frac{\partial\chi}{\partial a}
   -\frac1D\frac{d^2D}{da^2}
   +\frac2{D^2}\left(\frac{dD}{da}\right)^2\right]
  -\frac{f(a)}a\frac{\partial S_0}{\partial a}\frac{\partial S_2}{\partial a}\nonumber\\
&-&\frac{\hbar^2}4\left(\frac{f(a)}{a^2}-\frac1a\frac{df}{da}\right)
   \left(\frac1{\chi}\frac{\partial\chi}{\partial a}-\frac1D\frac{dD}{da}\right)
  +\frac i2\frac{f(a)}a\sum_{\bf k}\lambda_{\bf k}b^*_{\bf k}\frac{\partial S_2}{\partial b^*_{\bf k}},
\end{eqnarray}
the equation for $\sigma_2$, which is the analogue of (\ref{sigma2}):
\begin{equation}
\label{sigma2_EPS}
\frac{f(a)}a\frac{d\sigma_0}{da}\frac{d\sigma_2}{da}
 -\frac{\hbar^2}{4Da}\frac{dD}{da}\left(\frac{f(a)}a-\frac{df}{da}\right)
 +\hbar^2\frac{f(a)}{2a}\left[\frac1D\frac{d^2D}{da^2}-\frac2{D^2}\left(\frac{dD}{da}\right)^2\right]=0,
\end{equation}
and the equation for $\eta_2$, which is the analogue of (\ref{eta2}):
\begin{equation}
\label{eta2_EPS}
-\frac{\partial\eta_2}{\partial t}
=-\frac{f(a)}a\frac{d\sigma_0}{da}\frac{\partial\eta_2}{\partial a}
 +\hbar^2\frac{f(a)}{2a}\left(\frac1{\chi}\frac{\partial^2\chi}{\partial a^2}
 -\frac2{D\chi}\frac{dD}{da}\frac{\partial\chi}{\partial a}\right)
 -\frac{\hbar^2}{4a\chi}\left(\frac{f(a)}a-\frac{df}{da}\right)\frac{\partial\chi}{\partial a}
 +\frac i2\frac{f(a)}{a}\sum_{\bf k}\lambda_{\bf k}b^*_{\bf k}\frac{\partial\eta_2}{\partial b^*_{\bf k}}.
\end{equation}

In the extended phase space approach, as well as in the approach of Montani and his collaborators, one does not need to construct a time derivative artificially, since, thanks to introducing a reference frame, the time derivative is already present in the formalism. Therefore, one does not need to insert $D(a)$ in the definition of the matter wave function. In the extended phase space, we have a choice: to take the function $D(a)$ equal to $a^{\frac14}$ (see (\ref{D_a})) to reproduce the results of the approach of Kiefer and his collaborators, or to take it simply equal to 1.

In the first case, choosing also $f(a)=1$, we can see that the equation (\ref{sigma2_EPS}) is reduced to (\ref{sigma2}). After introducing the function (\ref{xi_fun}), the first term in the right-hand side of (\ref{eta2_EPS}) can be rewritten as was done in (\ref{t-d-tr}) and, keeping in mind (\ref{der-chi-xi}), we conclude that, in general, this term gives a quantum gravitational correction of the order
${\cal O}\left(\displaystyle\frac1M\right)$. To avoid the correction, one can make the assumption that $\eta_2$ does not depend on $a$, but only on scalar field variables. Then, repeating the steps similar to those described in (\ref{H_M_xi}) -- (\ref{xi-fin}) we come to the equation
\begin{equation}
\label{xi_fin_EPS}
i\hbar\frac{\partial\xi}{\partial t}
 =H_m\xi
 +\frac{\hbar^2}{2Ma}\left(\frac{\partial^2\xi}{\partial a^2}
  -\frac1a\frac{\partial\xi}{\partial a}\right).
\end{equation}

This equation formally coincides with (\ref{xi-fin}), except that the time $\tau$ in (\ref{xi-fin}) is a parameter along a classical trajectory, while $t$ in (\ref{xi_fin_EPS}) is the time in a chosen reference frame. There is no contradiction between these two times. If classical spacetime exists, which is implied, a reference frame can be chosen in such a way that these two times are in agreement or even coincide.

In the second case, we put $D(a)=1$ and keep $f(a)$ arbitrary. Now the equation (\ref{M-1-D-EPS}) looks like
\begin{equation}
\label{M-1-D1-EPS}
-\frac{\partial S_2}{\partial t}
 =\hbar^2\frac{f(a)}{2a\chi}\frac{\partial^2\chi}{\partial a^2}
  -\frac{f(a)}a\frac{\partial S_0}{\partial a}\frac{\partial S_2}{\partial a}
  -\frac{\hbar^2}{4a\chi}\left(\frac{f(a)}a-\frac{df}{da}\right)\frac{\partial\chi}{\partial a}
  +\frac i2\frac{f(a)}a\sum_{\bf k}\lambda_{\bf k}b^*_{\bf k}\frac{\partial S_2}{\partial b^*_{\bf k}}.
\end{equation}
Eq. (\ref{sigma2_EPS}) yields $\displaystyle\frac{d\sigma_2}{da}=0$, so that $\sigma_2$ is a constant which can be taken equal to zero. Again, to avoid the second term in the right-hand side of (\ref{M-1-D1-EPS}), we can choose $S_0=\eta_2(b^*_{\bf k})$. The equation for $\eta_2$ is reduced to
\begin{equation}
\label{eta2-D1_EPS}
-\frac{\partial\eta_2}{\partial t}
 =\hbar^2\frac{f(a)}{2a\chi}\frac{\partial^2\chi}{\partial a^2}
 -\frac{\hbar^2}{4a\chi}\left(\frac{f(a)}a-\frac{df}{da}\right)\frac{\partial\chi}{\partial a}
 +\frac i2\frac{f(a)}{a}\sum_{\bf k}\lambda_{\bf k}b^*_{\bf k}\frac{\partial\eta_2}{\partial b^*_{\bf k}}.
\end{equation}
Making use of the Hamiltonian $H_{mat}$ (\ref{Ham_mat_EPS}) instead of $H_m$, we get the following equation for $\xi$ (\ref{xi_fun}):
\begin{equation}
\label{xi_D1}
i\hbar\frac{\partial\xi}{\partial t}
 =H_{mat}\xi
  +\hbar^2\frac{f(a)}{2Ma}\frac{\partial^2\xi}{\partial a^2}
  -\frac{\hbar^2}{4Ma}\left(\frac{f(a)}a-\frac{df}{da}\right)\frac{\partial\xi}{\partial a}.
\end{equation}

In fact, this equation differs from (\ref{xi-fin}) and (\ref{xi_fin_EPS}). Even if $f(a)=1$, the equations do not coincide. In principle, since quantum gravitational corrections are different in various approaches, it could give a little hope to discriminate between the approaches if the accuracy of observations enables one to compare theoretical predictions with observational data.

\section{Discussion and conclusions}
In this paper, we have analysed three different approaches to taking the semiclassical limit of quantum gravity. All of them are based on rather arbitrary additional assumptions. The approach of Kiefer and his collaborators seems to be more grounded what concerns the Schr\"odinger equation with quantum-gravitational corrections. However, this approach, based on the Born -- Oppenheimer approximation for gravity, was recently criticized in \cite{CC}. It was noticed that even when the Born -- Oppenheimer approximation is applied to molecules, it is tacitly supposed that there exist some timescales. Indeed, to speak about slow or fast moving particles, one needs to have these timescales that cannot be explained just by the difference in masses of the particles. The same is true when one speaks that gravity changes slowly and the scalar field changes faster. Therefore, one has been caught in a vicious circle: one needs spacetime to explain the appearance of spacetime. The construction of the time operator from the right-hand side of (\ref{time_oper}) may be considered as an elegant mathematical trick, but it gives us no idea what had happened at the physical level. If time is a parameter along a trajectory in the configurational space, a classical spacetime exists. Hence, one can introduce a reference frame in this classical (macroscopic) spacetime and continue coordinate lines into the Very Early Universe, a region of the Planckian size.

Introducing a reference frame is supposed in the two other approaches. However, in the approach of Montani and his collaborations there are other assumptions. In particular, gravity and matter are assumed to be separated that is determined by the form of the action (\ref{MM_ass}). Moreover, most of the obtained corrections are an order of magnitude lower than the same corrections in the approach of Kiefer and his collaborators. The calculations in \cite{MM,GMMN} are based on the canonical formalism, the operator ordering is not fixed, and the results seem to be quite formal. For that, the main advantage of this approach that the corrections to the Schr\"odinger equation are Hermitian, rises doubts.

We have demonstrated, that, in the extended phase space approach, we can reproduce the results of Kiefer and his collaborators under special conditions. In general, we have obtained a gauge-depended Schr\"odinger equation for matter fields that is natural for this approach. But the first term in the right-hand side of (\ref{eta2_EPS}), which gives rise to the time derivative in the framework of the Wheeler -- DeWitt approach, seems to be redundant in our case. To avoid this term, we also have to make the assumption about the second-order contribution to expansion of the action,
$S_2=\eta_2\left(b^*_{\bf k}\right)$, i.e. $S_2$ depends only on scalar field variables.

As was noted in \cite{KP}, the correction terms enable one to calculate quantum-gravitational corrections to the CMB power spectrum that could be observable in principle, but too small to be observable in the present time. On the other hand, we need more reasonable theoretical predictions.

The problem of time in quantum gravity returns us to the question, what equation is more fundamental, the Wheeler -- DeWitt equation or the Schr\"odinger equation \cite{Shest2}?

\end{document}